\documentclass[10pt]{article}
\usepackage{amsfonts}
\usepackage{fullpage,graphicx,subfigure,mathdots,mathpazo,color}
\usepackage{amsmath,amscd,tikz,mathrsfs,cite}
\usepackage[normalem]{ulem}
\usepackage{amsmath}
\usepackage{setspace}
\usepackage{epsfig,amsmath,graphicx,amssymb,overpic}
\usepackage{bm}
\usepackage{graphicx}
\usepackage{subfigure, xcolor}
\usepackage{ntheorem}
\usepackage{listings,diagbox}
\usepackage{color}
\usepackage{epsfig,amsmath,graphicx,amssymb,overpic}
\usepackage{booktabs}
\usepackage{diagbox}
\usepackage{graphicx}
\usepackage{dcolumn}
\usepackage{bm}
\usepackage{graphicx}
\usepackage{subfigure}
\usepackage{epsfig,amsmath,graphicx,amssymb,overpic,cite}
\usepackage{makecell}
\usepackage{fullpage,graphicx,subfigure,mathdots,mathpazo,color}
\usepackage{amsmath,amscd,tikz,mathrsfs,cite}
\usepackage[normalem]{ulem}
\usepackage{amsmath}
\usepackage{setspace}
\usepackage{epsfig,amsmath,graphicx,amssymb,overpic}
\usepackage{bm}
\usepackage{graphicx}
\usepackage{subfigure, xcolor}
\usepackage{ntheorem}

\def\be{\begin{equation}}
\def\ee{\end{equation}}
\def\bee{\begin{eqnarray}}
\def\ene{\end{eqnarray}}
\def\bes{\begin{subequations}}
\def\ees{\end{subequations}}

\newtheorem{remark}{Remark}

\allowdisplaybreaks[4]

\begin{document}

\baselineskip=13pt \renewcommand {\thefootnote}{\dag}
\renewcommand
{\thefootnote}{\ddag} \renewcommand {\thefootnote}{ }

\pagestyle{plain}




\begin{flushleft}
\baselineskip=16pt \leftline{} \vspace{-.3in} {\Large \textbf{Formations and
dynamics of two-dimensional spinning asymmetric quantum droplets controlled
by a $\mathcal{PT}$-symmetric potential}} \\[0.2in]

Jin Song$^{1,2}$,\,\, Zhenya Yan$^{\mathrm{1,2,a)}}$, and Boris A. Malomed$%
^{3,4}$ 
\\[0.05in]
{\small $^1$KLMM, Academy of Mathematics and Systems Science, Chinese
Academy of Sciences, Beijing 100190, China \newline
$^2$School of Mathematical Sciences, University of Chinese Academy of
Sciences, Beijing 100049, China \newline
$^{3}$Department of Physical Electronics, School of Electrical Engineering,
Faculty of Engineering, Tel Aviv University, Tel Aviv 69978, Israel \newline
$^{4}$Instituto de Alta Investigaci\'{o}n, Universidad de Tarapac\'{a},
Casilla 7D, Arica 1000000, Chile \newline
$^{\mathrm{a)}}$\textbf{Author to whom correspondence should be addressed:}
zyyan@mmrc.iss.ac.cn } \newline
\end{flushleft}

{\baselineskip=13pt }


\vspace{-0.01in}


\noindent \textbf{Abstract:}\thinspace\ In this paper, vortex solitons are
produced for a variety of 2D spinning quantum droplets (QDs) in a $\mathcal{%
PT}$-symmetric potential, modeled by the amended Gross-Pitaevskii equation
with Lee-Huang-Yang corrections. In particular, exact QD states are obtained
under certain parameter constraints, providing a guide to finding the
respective generic family. In a parameter region of the unbroken $\mathcal{PT%
}$ symmetry, different families of QDs originating from the linear modes are
obtained in the form of multipolar and vortex droplets at low and high
values of the norm, respectively, and their stability is investigated. In
the spinning regime, QDs become asymmetric above a critical rotation
frequency, most of them being stable. The effect of the $\mathcal{PT}$%
-symmetric potential on the spinning and nonspinning QDs is explored by varying the strength
of the gain-loss distribution. Generally, spinning QDs trapped in the $%
\mathcal{PT}$-symmetric potential exhibit asymmetry due to the energy flow
affected by the interplay of the gain-loss distribution and rotation.
Finally, interactions between spinning or nonspinning QDs are explored, exhibiting elastic
collisions under certain conditions.



\vspace{-0.05in} 

\vspace{0.12in}



\vspace{0.3in}

\baselineskip=13pt

\textbf{As a new species of liquids, quantum droplets (QDs), originating
from the delicate balance between the mutual attraction and self-repulsion
in two-component Bose-Einstein condensates (BECs), attract steadily growing
interest in studies of ultracold atoms and superfluids, since the prediction
of QDs by Petrov \cite{Petrov2015}. They are stabilized against the critical
or supercritical collapse by the Lee-Huang-Yang (LHY) correction to
mean-field interactions, making it possible to observe stable QDs, such as
vortex QDs, multiple droplets, QD clusters, and crystals \cite{book}. On the
other hand, experimental realization of complex $\mathcal{PT}$ potentials in
BECs is feasible and many ideas for the respective experimental design have
been proposed recently. Especially, as concerns a variety of QD patterns
with a complex 2D arrangement, such as vortex droplets and multiple QDs and
QDs clusters, it is important to consider effects of the gain-loss
distribution in the $\mathcal{PT}$-symmetric potential on their structure.
In this paper, a rich variety of spinning QDs are found trapped in a $%
\mathcal{PT}$-symmetric potential, in the form of multipolar and vortex
droplets at low and high values of the norm, respectively, and their
stability is investigated. In the spinning regime, QDs become asymmetric
above a critical value of the rotation frequency, most of them being stable.
Furthermore, it is found that spinning QDs trapped in a $\mathcal{PT}$%
-symmetric potential exhibit asymmetry due to the energy flow affected by
the interplay of the gain-loss distribution and rotation. Finally,
interactions between spinning or nonspinning QDs are investigated which display their
elastic collisions under certain conditions. }

\section{Introduction}

Quantum droplets (QDs), as a new species of quantum matter, were proposed by
Petrov and Astrakharchik~\cite{Petrov2015,Petrov2016} for binary
Bose-Einstein condensates (BECs), taking into regard the self-repulsive
Lee-Huang-Yang (LHY) correction~to the Gross-Pitaevskii (GP) equations of
the mean-field (MF)\ theory, induced by quantum fluctuations \cite{LHY}.
Still earlier, it was proposed to create stable self-trapped states similar
to QDs, using not the LHY effect but three-body interactions \cite{Bulgac}.
Currently, the formation and dynamics of QDs are in the focus of studies of
superfluids and ultracold gases~\cite{Chomaz2016,dipolar2,Pfau-review,review,book}.
The competition between the LHY correction and the residual MF interaction
(a weak imbalance between the intra-component repulsion and inter-component
attraction in the binary BEC) maintains a superfluid state whose density
cannot exceed a certain maximum, which makes it incompressible. This is the
reason why this quantum macroscopic state of matter is identified as a
fluid, and localized states filled by it are called droplets \cite{review}.
In comparison to classical fluids, QDs exist with extremely low densities at
extremely low temperatures, offering possibilities to observe complex
many-body effects \cite{rosi,prb}. QDs have been experimentally realized in
ingle-component dipolar bosonic gases \cite{dipolar1,Kadau2016,nat}, binary
Bose-Bose mixtures of two different atomic states in $^{39}$K \cite%
{Cabrera2018,Cheiney2018,Semeghini2018,exp-PRL2019,poo}, and in the
heteronuclear mixture of $^{41}$K and $^{87}$Rb atoms \cite{exp-PRR2019}. In
addition to QDs in the one-dimensional (1D) geometry~\cite%
{Petrov2016,1d-2,boris2020pra,1d-3,1d-4}, they have been studied in detail
in multidimensional settings, such as vortex QDs, multiple droplets, QD
clusters, and crystals~\cite%
{vo2d,Kartashov2019,boris2020,vo3d,vo,md,md1,cc1,cc2,dc,Dong1}. In
particular, the existence and stability of 1D holes and kinks and 2D
vortices nested in extended BECs were investigated in Ref.~\cite{cc1}, while
Refs.~\cite{cc2} and \cite{Maluckov}  studied modulational instability in
binary BECs under the action of quantum fluctuations. A crucially important
fact is that the LHY effect makes it possible to suppress the critical or
supercritical collapse predicted by the MF theory with intrinsic attraction,
thus securing the stability of the QDs in the 2D and 3D geometries.

External potentials are a necessary ingredient in the experimental
realization of QDs. The potentials can be also used as an efficient means
for shaping QDs and controlling their dynamics \cite%
{po,poo,po1,po2,Hu,Raymond}. In particular, dynamics of QDs in an external
harmonic-oscillator (HO)\ confinement was investigated in Ref. \cite{po1}.
Different types of 2D and 3D stable QDs persistently spinning in an
anharmonic potential were predicted too \cite{po2}.

Further, the complex potential $U({\mathbf{r}})$, subject to the constraint
\begin{equation}
U^{\ast }(\mathbf{r})=U(-\mathbf{r}),  \label{PT}
\end{equation}%
where $\ast $ stands for the complex conjugate, makes it possible to
introduce a promising extension of standard quantum mechanics with the
parity-time ($\mathcal{PT}$) symmetry, as proposed by Bender and Boettcher
in 1998 \cite{bender}. Non-Hermitian Hamiltonians including complex
potentials subject to constraint (\ref{PT}) admit fully real (physically
meaningful) spectra. The symmetry is realized in this case with parity
operator $\mathcal{P}$ and time-reversal one $\mathcal{T}$ defined as
\begin{equation}
\mathcal{P}:{\mathbf{r}}\rightarrow -{\mathbf{r}};\quad ~\mathcal{T}%
:i\rightarrow -i,\,\, t\rightarrow -t.  \label{PT2}
\end{equation}%
A variety of $\mathcal{PT}$-symmetric potentials, such as the Scarf-II
potential \cite{scarf1,scarf2}, Gaussian \cite{g1,g2}, and HO \cite{har1,har2,har3}
 potentials, have been shown to support stable soliton solutions in
various nonlinear models \cite{rmp-16}. A number of novel wave phenomena
related to the $\mathcal{PT}$-symmetry have been observed experimentally in
optics \cite{25,26,27,Peschel}, where the roles of the real and imaginary
parts of the potential are played, respectively, by a spatially symmetric
modulated profile of the refractive index and a spatially antisymmetric
distribution of the local gain and loss.

Realization of complex potentials subject to constraint (\ref{PT}) is
feasible in other physical settings too~\cite{rmp-16}, including BEC \cite%
{Cartarius}. In particular, a quantum-mechanical implementation of the
concept was proposed in Ref. \cite{ex1}, where $\mathcal{PT}$-symmetric
currents are represented by an accelerating BEC in a titled optical lattice.
It was shown that, by setting suitable initial currents between neighboring
lattice sites, it is possible to create $\mathcal{PT}$-symmetric states in a
two-mode subsystem embedded into the lattice~\cite{ex2}. Spontaneous breaking of the $%
\mathcal{PT}$ symmetry in a dual-core trap and dynamics of 1D droplets
trapped in an optical-lattice potential were investigated too \cite%
{Zezyulin,ep1,ep2,ep3}. The dynamics of one- and three-dimensional QDs in $%
\mathcal{PT}$-symmetric HO-Gaussian (HOG) potentials was also studied \cite%
{songqd}.

As concerns a variety of QD patterns with a complex 2D arrangement, such as
vortex droplets, multiple QDs and QD clusters \cite{vo2d,md1,po}, it is
relevant to consider effects of the gain-loss distribution in the $\mathcal{%
PT}$-symmetric potential on their structure. In this context, we introduce
generalized $\mathcal{PT}$-symmetric HOG potentials into the 2D GP equation
and investigate the dynamics of QDs representing ground and excited states.
Families of spinning QD sets and their stability are discussed in detail
too. The main results produced by this paper can be summarized as follows:

\begin{itemize}
\item {} The generalized complex $\mathcal{PT}$-symmetric HOG potential is
added to the 2D GP equations including the LHY correction. Exact QDs
solutions are found under certain conditions imposed on the system's
parameters, which is a guide helping to construct the respective generic QD
family. Different families originating from linear modes are obtained. They
appear as multipolar droplets at a low norm and vortex droplets at large
norms, and their stability is investigated in detail.

\item {} In the regime of rotation, the formation of QDs and their stability
is explored in detail. It is found that QDs acquire an asymmetric form above
a critical rotation frequency, remaining stable states.

\item {} The effect of the $\mathcal{PT}$-symmetric potential on the
spinning QDs is discussed by adjusting the strength of the gain-loss
distribution (the imaginary part of the complex potential). The asymmetric
shape is caused by the asymmetric energy flow resulting from an imbalance in
the gain-loss distribution acting on the rotating patterns.

\item {} Collisions between the spinning QDs are explored in the presence of
the $\mathcal{PT}$-symmetric HOG potential. It is found that collisions are
elastic, under certain conditions.
\end{itemize}

The remainder of the paper is organized as follows. In Sec. 2, the GP
equation including the LHY correction and $\mathcal{PT}$-symmetric
potentials is introduced. In Sec. 3, we introduce the $\mathcal{PT}$%
-symmetric HOG potential, and analyze the real spectra of the corresponding
non-Hermitian Hamiltonian. Particular 2D exact solutions for nonspinning QDs
are given. Different families of QDs originating from linear modes are
obtained, and their stability is investigated in detail. In Sec. 4, 2D
spinning multiple-component QD structures are produced with different
shapes, depending on the angular velocity. Furthermore, to understand the
effect of the $\mathcal{PT}$-symmetric potential on the spinning or nonspinning QDs,
we adjust the strength of the gain-loss distribution to investigate
the QD\ dynamics in Sec. 5. In Sec. 6, interactions between spinning or nonspining QDs are
addressed. The paper is concluded by Sec. 7.

\section{2D amended GP equation with the rotation, LHY correction, and $%
\mathcal{PT}$ potential}


\quad \textit{The $\mathcal{PT}$-symmetric models for 2D spinning QDs}.---In
the binary BECs with two mutually symmetric components trapped in a 2D $%
\mathcal{PT}$-symmetric potential, the underlying GP equation with the LHY
correction (represented by the cubic nonlinearity with the additional
logarithmic factor) takes the form of~\cite{Petrov2015,Petrov2016,review,Cabrera2018,boris2020}:
\begin{equation}
\begin{array}{rl}
i\partial _{T}\Psi = & \!\!\!-\dfrac{1}{2}\nabla _{\mathbf{R}}^{2}\Psi +%
\dfrac{8\pi }{\ln ^{2}(a_{\uparrow \downarrow }/a)}\ln \left( \dfrac{|\Psi
|^{2}}{\sqrt{e}n_{0}}\right) |\Psi |^{2}\Psi +[\mathrm{U}_{R}(\mathbf{R})+i%
\mathrm{U}_{I}(\mathbf{R})]\Psi ,%
\end{array}
\label{3d0}
\end{equation}%
where the complex wave function $\Psi =\Psi (\mathbf{R},T)$ represents equal
wave functions $\Psi _{1}=\Psi _{2}\equiv \Psi /\sqrt{2}$ of two components $%
(\Psi _{1},\Psi _{2})$ of the binary BEC,
\begin{equation}
\mathbf{R}=(X,Y)=\mathcal{R}/r_{0}\,\,(\mathcal{R}=(\mathcal{X},\mathcal{Y})),\quad  T=\tau /t_{0}  \label{RT}
\end{equation}
stand for the 2D scaled coordinates and time, respectively, $r_{0}$ is the
characteristic spatial scale, which defines the characteristic energy and
time scales,
\begin{equation}
\varepsilon _{0}=\hbar ^{2}/mr_{0}^{2}, \quad  t_{0}=\hbar /\varepsilon _{0},
\label{t0}
\end{equation}
$m$ is the atomic mass, $\nabla _{\mathbf{R}}^{2}=\partial ^{2}/\partial
X^{2}+\partial ^{2}/\partial Y^{2}$ denotes the 2D Laplacian, $\mathrm{U}%
_{R}(\mathbf{R})+i\mathrm{U}_{I}(\mathbf{R})=\mathcal{U}(\mathcal{R}%
)/\varepsilon _{0}$ is a scaled $\mathcal{PT}$-symmetric potential with $%
\mathrm{U}_{R}(\mathbf{R})$ being a real-valued potential, and $\mathrm{U}%
_{I}(\mathbf{R})$ denoting the gain-loss distribution. Further, the
Feshbach-resonance technique, acting by means of external magnetic field $B$,
 allows one to tune the intra- and inter-state scattering lengths of atomic
collisions, $a_{\uparrow \uparrow },\,a_{\downarrow \downarrow }>0$, $%
a\equiv \sqrt{a_{\uparrow \uparrow }a_{\downarrow \downarrow }}$, and $%
a_{\uparrow \downarrow }<0$. Further, $\delta a=a_{\uparrow \downarrow }+%
\sqrt{a_{\uparrow \uparrow }a_{\downarrow \downarrow }}$ with $|\delta a|\ll
a$ denotes the small imbalance between the intracomponent repulsion and
intercomponent attraction~\cite{boris2020,Cheiney2018}, and $n_{0}=2\pi \exp
(-2\gamma -3/2)(aa_{\uparrow \downarrow })^{-1}\ln (a_{\uparrow \downarrow
}/a)$ is the equilibrium density of each component ($\gamma \approx 0.5772$
is the Euler's constant). For real BEC, we consider $^{39}$K atoms tightly
confined by a transversal external potential. The experiment used a
sufficiently large magnetic field, $B\approx 55.5$ G, to fix $\delta
a\approx -3.2a_{0}<0$ by means of the Feshbach resonance, where $a_{0}$ is
the Bohr radius \cite{Cabrera2018,Cheiney2018,Semeghini2018}. The scaling
transform, $\Psi =\varrho _{0}\psi ,\,\mathbf{R}=(\varrho _{0}/2)^{1/2}%
\mathbf{r}$ and $T=\varrho _{0}t$ with $\varrho _{0}=(\sqrt{e}n_{0})^{1/2}$
and $\mathbf{r}=(x,y)$, casts Eq.~(\ref{3d0}) in the normalized form,
\begin{equation}
i\partial _{t}\psi =-\nabla _{\mathbf{r}}^{2}\psi +\sigma \ln (|\psi
|^{2})|\psi |^{2}\psi +[V(\mathbf{r})+iW(\mathbf{r})]\psi ,  \label{3d}
\end{equation}%
where $\sigma =8\pi (\sqrt{e}n_{0})^{3/2}/\ln ^{2}(a_{\uparrow \downarrow
}/a)>0$ represents the strength of the logarithmic LHY contribution to
nonlinearity. The complex $\mathcal{PT}$-symmetric potential $U(\mathbf{r}%
)\equiv V({\mathbf{r}})+iW({\mathbf{r}})=\varrho _{0}(\mathrm{U}_{R}(\mathbf{%
R})+i\mathrm{U}_{I}(\mathbf{R}))$ satisfies constraint (\ref{PT}), securing
the invariance of Eq.~(\ref{3d}) with respect to the $\mathcal{PT}$
transformation (\ref{PT2}). Equation~(\ref{3d}) can be written in the
Hamiltonian form, $i\partial _{t}\psi =\delta \mathcal{E}(\psi ,\psi ^{\ast
})/\delta \psi ^{\ast }$, with quasi-energy
\begin{equation}
\mathcal{E}(\psi ,\psi ^{\ast })=\int \!\!\int_{\mathbb{R}^{2}}\left[
|\nabla _{\mathbf{r}}\psi |^{2}+[V({\mathbf{r}})+iW({\mathbf{r}})]|\psi
|^{2}+\frac{\sigma }{4}(2\ln (|\psi |^{2})-1)|\psi |^{4}\right] \mathrm{d}%
^{2}{\mathbf{r}}.  \label{E}
\end{equation}%
The quasi-energy and the norm of the wave function,
\begin{equation}
N=\int \!\!\int_{\mathbb{R}^{2}}|\psi (\mathbf{r},t)|^{2}\mathrm{d}^{2}{%
\mathbf{r}},  \label{N}
\end{equation}%
are dynamical invariants of Eq. (\ref{3d}) only in the case when the density
field is spatially even, $\left\vert \psi (-\mathbf{r},t)\right\vert
=\left\vert \psi (\mathbf{r},t)\right\vert $, or if the imaginary part of
the potential is absent.

\begin{remark}
The nonlinear term in the 3D GP equation is a combination of the usual MF
cubic nonlinearity and a quartic defocusing LHY term. To derive the 2D
equation (\ref{3d}), it is assumed that a strong confining potential is
applied in the $z$ direction, with confinement width $a_{\bot }$ which is
much smaller than the healing length corresponding to the equilibrium
density, $\zeta =(32\sqrt{2}/3\pi )(a/|\delta a|)^{3/2}a)$. In this case,
the dimensional reduction, 3D $\rightarrow $ 2D, transforms the
cubic-quartic nonlinearity into the cubic-logarithmic term, as it is written
in Eq. (\ref{3d}). In the opposite case, with $a_{\perp }\gg \zeta $, the
effective 2D GP equation keeps the same cubic-quartic form as in 3D \cite%
{boris2020}. To guarantee the validity of the cubic-logarithmic
nonlinearity, the necessary condition for the confinement size is $a_{\bot }
\ll \zeta \sim 30$ nm \cite{Cabrera2018,Cheiney2018,Semeghini2018}, and the
actual value of $a_{\bot }$ used in the experiment is $\sim0.6$ $\mu\mathrm{m%
}$.
\end{remark}

To seek for solutions for spinning QDs, Eq. (\ref{3d}) is rewritten in terms
of spinning coordinates, $x^{\prime }=x\cos (\omega t)+y\sin (\omega t)$, $%
y^{\prime }=y\cos (\omega t)-x\sin (\omega t)$ with angular velocity $\omega
$:
\begin{equation}
i\partial_t\psi=-\nabla ^{2}\psi +i\omega \widehat{\mathbf{r}}\cdot \nabla _{%
\mathbf{r}}\psi +\sigma \ln (|\psi |^{2})|\psi |^{2}\psi +[V(\mathbf{r})+iW(%
\mathbf{r})]\psi ,  \label{3dr}
\end{equation}%
where $\widehat{\mathbf{r}}=(y,-x)$ and $\cdot $ stands for the scalar
product. The spinning term $i\omega \widehat{\mathbf{r}}\cdot \nabla _{%
\mathbf{r}}\psi $ in Eq. (\ref{3dr}) can be counterbalanced by the complex $%
\mathcal{PT}$ potential $V(\mathbf{r})+iW(\mathbf{r})$. Note that, for a
fixed nonlinearity coefficient $\sigma>0$, the increase of the local density
$|\psi (\mathbf{r},t)|^{2}$ from small ($<1$) to large ($>1$) values leads
to the change of the sign of the logarithmic factor in nonlinearity of Eq.~(%
\ref{3dr}), i.e., the switching from self-attraction (focusing) to repulsion
(defocusing).



\vspace{0.1in} \textit{Stationary solutions and the stability analysis.}%
---First, we focus on spinning QDs produced by Eq.~(\ref{3dr}) in the form
of stationary solutions,
\begin{equation}
\psi ({\mathbf{r}},t)=\phi (\mathbf{r})e^{-i\mu t},  \label{psiphi}
\end{equation}%
where $\mu $ is the chemical potential, and the stationary wave function $%
\phi (\mathbf{r})$ is localized, vanishing at $|{\mathbf{r}}|\rightarrow
\infty $. It follows from Eqs. (\ref{3dr}) and (\ref{psiphi}) that the
localized eigenmode $\phi (\mathbf{r})$ obeys the following complex
nonlinear stationary equation:
\begin{equation}
\mu \phi =-\nabla ^{2}\phi +i\omega \widehat{\mathbf{r}}\cdot \nabla _{%
\mathbf{r}}\phi +\sigma \ln (|\phi |^{2})|\phi |^{2}\phi +[V(\mathbf{r})+iW(%
\mathbf{r})]\phi .  \label{3ds}
\end{equation}%
Below, we produce exact analytical solutions of Eq. (\ref{3ds}), which exist
for the possible specific conditions. Generic solutions of Eq. (\ref{3ds})
for QDs with zero-boundary conditions at $|{\mathbf{r}}|\rightarrow \infty $
can produced in a numerical form by means of techniques such as the
squared-operator iterations \cite{49}, Newton-conjugate-gradient method~\cite%
{50}, and spectral renormalization~\cite{51}. In this paper, we mainly use
the Newton-conjugate-gradient method, since it converge much faster than the
other existing iteration methods, often by orders of magnitude.

Stability of spinning QDs is investigated numerically by adding a small
perturbation to the stationary solutions of Eq.~(\ref{3dr}), as
\begin{equation}
\psi (\mathbf{r},t)=\big\{\phi (\mathbf{r})+\nu \lbrack F(\mathbf{r}%
)e^{-i\varepsilon t}+G^{\ast }(\mathbf{r})e^{i\varepsilon ^{\ast }t}]\big\}%
e^{-i\mu t},  \label{perturb}
\end{equation}%
where $\nu \ll 1$, $F(\mathbf{r})$ and $G(\mathbf{r})$, and $\varepsilon $
stand for, respectively, the infinitesimal amplitude, eigenfunctions, and
instability growth rate of the perturbation. Substituting the perturbed
solution (\ref{perturb}) in Eq.~(\ref{3dr}) and linearizing it with respect
to $\nu $ can lead to the Bogoliubov-de Gennes equations,
\begin{equation}
\left(
\begin{matrix}
L_{1} & L_{2}\vspace{0.05in} \\
-L_{2}^{\ast } & -L_{1}^{\ast }%
\end{matrix}%
\right) \left(
\begin{matrix}
F(\mathbf{r})\vspace{0.1in} \\
G(\mathbf{r})%
\end{matrix}%
\right) =\varepsilon \left(
\begin{matrix}
F(\mathbf{r})\vspace{0.1in} \\
G(\mathbf{r})%
\end{matrix}%
\right) ,  \label{linear3d}
\end{equation}%
where the linear operators are
\begin{eqnarray}
{\nonumber} L_{1}=-\nabla_{\mathbf{r}}^{2}+i\omega \widehat{\mathbf{r}}%
\cdot \nabla _{\mathbf{r}}+(V+iW)+\sigma |\phi |^{2}[2\ln (|\phi
|^{2})+1]-\mu,\quad L_{2}=\sigma \phi ^{2}[\ln (|\phi |^{2})+1].
\end{eqnarray}
The underlying stationary solutions are unstable if the spectrum of
eigenvalues $\varepsilon $ contains ones with nonvanishing imaginary parts.
The spectrum $\varepsilon $ can be produced, solving Eq. (\ref{linear3d})
numerically by means of the Fourier collocation method \cite{52}. Then, the
predicted (in)stability is verified by direct simulations the perturbed
evolution in the framework of Eq. (\ref{3dr}).

\section{2D linear and nonlinear regimes for nonspinning QDs}

\subsection{Phase transitions for the $\mathcal{PT}$-symmetry}

\begin{figure}[t]
\centering
{\scalebox{0.75}[0.75]{\includegraphics{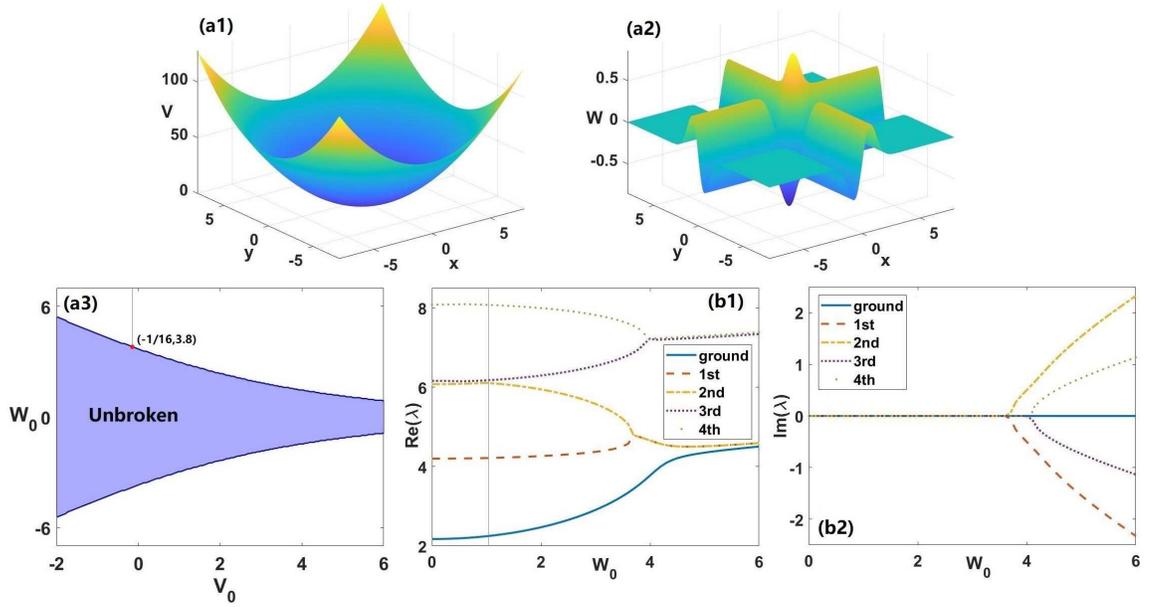}}}\hspace{-0.35in} \vspace{%
0.1in}
\caption{{\protect\small (a1,a2) Real and imaginary parts of the 2D $%
\mathcal{PT}$-HOG potential (\protect\ref{hg}) with $%
(V_{0},V_{1},W_{0})=(-1/16,1,1)$. (a3) The blue (white) shaded area shows
the area of unbroken (broken) $\mathcal{PT}$} symmetry, as produced by the
numerical solution of the {\protect\small linear spectral problem (\protect
\ref{ls}) with the $\mathcal{PT}$-HOG potential (\protect\ref{hg}) as $%
V_1=1$. (b1,b2) Real and imaginary parts of the lowest energy eigenvalues $%
\protect\lambda $ as functions of $W_{0}$ at fixed $V_{0}=-1/16$ and $V_1=1$%
. }}
\label{pt}
\end{figure}

In this subsection, we first introduce a generalized 2D $\mathcal{PT}$%
-symmetric HOG ($\mathcal{PT}$-HOG) potential
\begin{equation}
\mathcal{U}(\mathcal{R})=m\Omega ^{2}\left\{ \left[ \mathcal{R}^{2}(\kappa
_{0}+\kappa _{1}e^{-\mathcal{R}^{2}/d^{2}})+\kappa _{2}(e^{-2(\mathcal{X}%
/d)^{2}}+e^{-2(\mathcal{Y}/d)^{2}})\right] +i\kappa _{3}\left( \frac{%
\mathcal{X}}{d}e^{-2(\mathcal{X}/d)^{2}}+\frac{\mathcal{Y}}{d}e^{-2(\mathcal{%
Y}/d)^{2}}\right) \right\} ,  \label{hU}
\end{equation}%
where $\Omega $ is the trapping frequency, $d=r_{0}\sqrt{\varrho _{0}/2}$ is
the oscillator length, $\mathcal{R}^{2}=\mathcal{X}^{2}+\mathcal{Y}^{2}$,
and $\kappa _{j}$ ($j=0,1,2,3$) are free potential parameters. Then one gets
a dimensionless $\mathcal{PT}$-symmetric potential, according to the scale
transform defined by Eqs. (\ref{RT}) and (\ref{t0}), with the real and
imaginary parts being
\begin{equation}
V(\mathbf{r})=r^{2}\left( 1+V_{1}e^{-r^{2}}\right) +V_{0}\,\left(
e^{-2x^{2}}+e^{-2y^{2}}\right) ,\qquad W(\mathbf{r})=W_{0}\left(
xe^{-x^{2}}+ye^{-y^{2}}\right) ,  \label{hg}
\end{equation}%
where $r^{2}=x^{2}+y^{2}$, the coefficient in front of the HO potential is
set to be $1$ by taking $\kappa _{0}=\varepsilon _{0}/(m\Omega
^{2}d^{2}\varrho _{0})$, real parameters $V_{0}=\kappa _{2}m\Omega ^{2}$ and
$V_{1}=\kappa _{1}m\Omega ^{2}\varrho _{0}/(\varepsilon _{0}d^{2})$ modulate
the profile of the external potential $V(\mathbf{r})$, and real parameter $%
W_{0}=\kappa _{3}m\Omega ^{2}$ is the strength of gain-loss distribution $W(%
\mathbf{r})$ [see Figs. \ref{pt}(a1,a2)]. For the characteristic transverse
scale of $r_{0}\sim 0.5$ $\mathrm{\mu m}$, one gets energy $\varepsilon
_{0}\approx 6.9\times 10^{-31}$ J. The free parameters, $\kappa
_{j}\,(j=0,1,2,3)$, can be adjusted to generate the proper shape of the
potential. Then one concentrates on the basic problem of the $\mathcal{PT}$%
-symmetry breaking phenomena in the framework of the nonspinning ($\omega =0$%
) linearized equation (\ref{3ds}):
\begin{equation}
\mathcal{H}\Phi (\mathbf{r})=\lambda \Phi (\mathbf{r}),\quad \mathcal{H}%
=-\nabla _{\mathbf{r}}^{2}+V(\mathbf{r})+iW(\mathbf{r}),\quad  \label{ls}
\end{equation}%
where $\lambda $ and $\Phi (\mathbf{r})$ are the eigenvalue and localized
eigenfunction, respectively. The linear spectral problem (\ref{ls}) can be
solved numerically by dint of the Fourier spectral method \cite{52}. Then, $%
\Phi (\mathbf{r})$ will be used as the input for numerical solution of Eq.~(%
\ref{3ds}) in its full form, including the rotation and nonlinearity. In the
next section we produce a particular exact soliton solution of nonlinear
equation~(\ref{3ds}), which differs from previously known $\mathcal{PT}$%
-symmetric ones~\cite{chen-cnsns}. Actually, the availability of the exact
solution is an incentive for the investigation of the complex potential (\ref%
{hg}).

Boundaries between unbroken and broken $\mathcal{PT}$ symmetry threshold
curves are plotted in Fig. \ref{pt}(a3), as produced by the numerical
solution of the linear spectral problem (\ref{ls}) with the $\mathcal{PT}$%
-HOG potential (\ref{hg}). The solution is obtained by means of the Fourier
spectral method in the domain of $(V_{0},W_{0})\in \lbrack -2,6]\times
\lbrack -7,7]$ with fixed $V_{1}=1$. It is obvious that the increase of $%
V_{0}$ leads to gradual shrinkage of the region of the unbroken $\mathcal{PT}
$ symmetry. For a fixed value of $V_{0}$, there always exists a threshold
value of $\left\vert W_{0}\right\vert $, beyond which a phase transition
occurs, leading to appearance of complex spectra. i.e., breaking of the $%
\mathcal{PT}$ symmetry.

Additionally, the real and imaginary parts of few lowest eigenvalues are
displayed in Figs.~\ref{pt}(b1,b2) for $V_{0}=-1/16$ and $V_1=1$, which
shows that the $\mathcal{PT}$ symmetry gets broken, at critical points,
through the collision of the real eigenvalues corresponding to the first and
second excited state. When $W_{0}$ exceeds the critical value ($W_{0}=3.8$),
a pair of complex conjugate eigenvalues arises the spectrum. And it is found
that the unbroken-symmetry region will extend as a very narrow stripe to
indefinitely large values of $V_0$.

\subsection{2D nonspinning QDs and their stability in the $\mathcal{PT}$%
-symmetric potential}

In this subsection, we first exhibit exact nonlinear modes for Eq.~(\ref{3ds}%
) with $\omega =0$ (without rotation) 
in the presence of the $\mathcal{PT}$-HOG potential (\ref{hg}). Then, we
numerically produce families of generic solutions for the QDs, which include
the particular analytical solutions. These are multipolar droplets at low
norms $N$ and vortex ones at large $N$. Their stability is investigated in
detail.

\subsubsection{Exact solutions for 2D nonspinning QDs}

\quad \textit{2D stationary QDs.}---Equation~(\ref{3ds}) with the chemical
potential $\mu =2$ and $\omega =0$ admits the following exact solution, with
a nontrivial phase structure, while the local density is the same as in the
ground state of the 2D linear Schr\"{o}dinger equation with the HO potential
(i.e., $|\phi (\mathbf{r})|^{2}=\exp (-r^{2})$):
\begin{equation}
\phi (\mathbf{r})\!=\!\exp \!\left\{ -\frac{r^{2}}{2}-\frac{i\sqrt{\pi }W_{0}%
}{8}\left[ \mathrm{erf}\left( \frac{x}{2}\right) +\mathrm{erf}\left( \frac{y%
}{2}\right) \right] \right\} ,  \label{sol2d}
\end{equation}%
where the standard error function is defined as $\mathrm{erf}(x)=(2/\sqrt{%
\pi })\int_{0}^{x}e^{-\eta ^{2}}\mathrm{d}\eta $, the norm (\ref{N}) of the
solution being, obviously, $N=\pi $. The solution is a nongeneric one, as it
is valid only if parameters $V_{0,1}$ of the real part of potential (\ref{hg}%
) and nonlinearity coefficient $\sigma $ in Eq. (\ref{3ds}) are related by
the following constraints:
\begin{equation}
V_{0}=-W_{0}^{2}/16,\qquad V_{1}=\sigma .  \label{relation}
\end{equation}
In particular, when the imaginary part absents in the $\mathcal{PT}$-HOG
potential, i.e., $W_{0}=0$, Eq. (\ref{relation}) demonstrates that the
complex $\mathcal{PT}$-HOG potential (\ref{hg}) reduces to the real form,
\begin{equation}
V(\mathbf{r})=r^{2}\left[ 1+\sigma \exp(-r^{2}) \right] ,
\end{equation}%
and the corresponding exact solution (\ref{sol2d}) is identical to the
above-mentioned ground-state wave function of the 2D linear Schr\"{o}dinger
equation with the usual HO potential, $\phi (\mathbf{r})=\exp \left(
-r^{2}/2\right) $.


\begin{figure}[t]
\centering
{\scalebox{0.8}[0.8]{\includegraphics{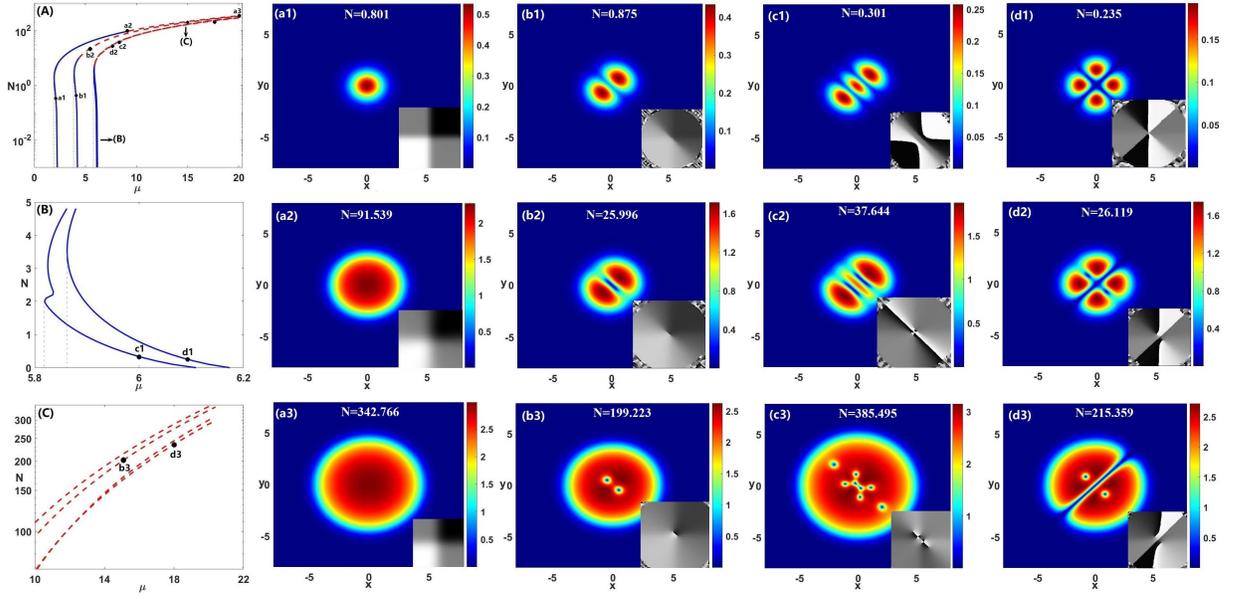}}} \vspace{0.12in}
\caption{{\protect\small (A) The dependence of norm $N$ on chemical
potential $\protect\mu $ for different families of droplet modes, as
produced by the full nonlinear equation (\protect\ref{3dr}) with $\protect%
\omega =0$ and $\protect\sigma =1$ (dashed: unstable; solid: stable). (B,C)
Zooms of the corresponding locations in (A). Points \textbf{a1, a2, a3,
b1,b2, b3, c1, c2, c3, d1, d2, d3} correspond to examples of QD modes
displayed in panels (a1)-(d3).} The density and phase structure of
one-component QDs: (a1) $\protect\mu =2$, (a2)  $\protect\mu =9$, and
(a3)$\protect\mu =20$. They correspond, respectively, to points \textbf{%
a1}, \textbf{a2}, and \textbf{a3} in Fig.~\protect\ref{stability}(A).
Examples of two-component QDs: (b1) $\protect\mu =4$, (b2) $\protect%
\mu =5.75$, and (b3) $\protect\mu =15$. They correspond, respectively,
to points \textbf{b1}, \textbf{b2}, and \textbf{b3} in Figs. \protect\ref%
{stability}(A,C). Examples of three-component QDs: (c1)  $\protect\mu =6$%
, (c2) $\protect\mu =8.2$, and (c3)  $\protect\mu =25$. They
correspond, respectively, to points \textbf{c1}, \textbf{c2}, and \textbf{c3}
in Figs. \protect\ref{stability}(A,B). Examples of four-component QDs: (d1)
 $\protect\mu =6.1$, (d2) $\protect\mu =7.5$, and (d3)  $\protect%
\mu =18$. They correspond, respectively, to points \textbf{d1}, \textbf{d2},
and \textbf{d3} in Figs. \protect\ref{stability}(A,B,C). Values of norm $N$
are indicated in each panel.}
\label{stability}
\end{figure}

\subsubsection{Numerical solutions for stationary QDs and their stability}

The exact solution (\ref{sol2d}) being available only under constraint (\ref%
{relation}) makes it necessary to find a family of generic solutions for
trapped states in a numerical form. To this end, Eq.~(\ref{3ds}) with $%
\omega =0$ was solved by means of the Newton-conjugate-gradient method.

First, we aim to find solutions of the linearized equation (\ref{3ds}) with $%
\sigma =0$, as nonlinear states bifurcate from the linear modes. In
particular, we produce the results for parameters
\begin{equation}
W_{0}=1,\qquad V_{0}=-1/16,  \label{1 and 1/16}
\end{equation}
which corresponds to the first relation in constraint (\ref{relation}), and
may adequately represent the generic case.

The linear spectra produced by the linearized equation (\ref{3ds}) with $%
V_{0}=-1/16$, fixed as per Eq. (\ref{1 and 1/16}) and varying strength of
the strength of the imaginary part of the potential are displayed in Fig.~%
\ref{pt}(b1). In particular, at $W_{0}=1$, taken as per Eq. (\ref{1 and 1/16}%
), the first four eigenvalues are $\lambda _{0}=2.2473,~\lambda
_{1}=4.216,~\lambda _{2}=6.1097,~\lambda _{3}=6.1739$, the spectra being
pure real.

\vspace{0.1in} \textit{Droplets with the one-component structure}.---$%
\mathcal{PT}$-symmetric droplets with the simplest structure originate from
the above-mentioned linear mode at $\lambda _{0}=2.2473$ with topological
charge $m=0$. The dependence between the norm and chemical potential $\mu $
for this solution family is displayed by the leftmost curve in Fig.~\ref%
{stability}(A). It is seen that the family exists at $\mu >1.9485$. At this
point, two different half-branches connect, the bottom and top ones. Their
stability is investigated by means of the numerical solutions of the
eigenvalue problem based on Eq.~(\ref{linear3d}). The solid blue and dashed
red solid segments of the $N(\mu )$ curves represent stable and unstable QD
states, respectively.

Examples of the densities and phase structures of QDs of the present types
are shown in Fig.~\ref{stability}(a1) for $\mu =2$ (a stable solution marked
\textbf{a1} on the bottom branch in Fig. \ref{stability}(A)),
Fig. \ref{stability}(a2) for $\mu =9$ (a solution marked \textbf{a2} on the
top branch in Fig. \ref{stability}(A), where it is located at the stability
boundary), and Fig. \ref{stability}(a3) for $\mu =20$ (an unstable solution
marked \textbf{a3} on the top branch in Fig. \ref{stability}(A)). Note that
the two latter modes feature a flat-top shape with a nearly uniform density
in the inner zone.

\vspace{0.1in} \textit{Droplets with the two-component structure}.---Dipole $%
\mathcal{PT}$-symmetric QDs composed of two separated fragments bifurcate
from the above-mentioned linear mode at $\lambda _{1}=4.216$, which
represents the first excited state of the linearized equation (\ref{3ds}).
The phase structure of these modes reveals that it carries topological
charge $m=1$. The respective QD family is represented by the middle curve in
Fig.~\ref{stability}(A)), which exists at $\mu \geq 3.90054$. At this point,
similar to the single-component QD family, the present one is composed of
two half-branches connecting at $\mu =3.90054$. An example of the density
and phase structure produced by the numerical solution for the stable
two-component QD at $\mu =4 $ (it corresponds to point \textbf{b1 }in Fig. %
\ref{stability}(A)), is shown in Fig.~\ref{stability}(b1). With the increase
of the norm the two QD components attract each other and gradually merge
into vortex dipoles, as shown in Fig. \ref{stability}(b2) for $\mu =5.75$.
At still larger values of $N$, the two components fuse into a single vortex
dipole carried by the flat-top background, with a phase singularity at the
center, as shown for $\mu =15$ in Fig.~\ref{stability}(b3).

\begin{figure}[!t]
\centering
{\scalebox{0.82}[0.85]{\includegraphics{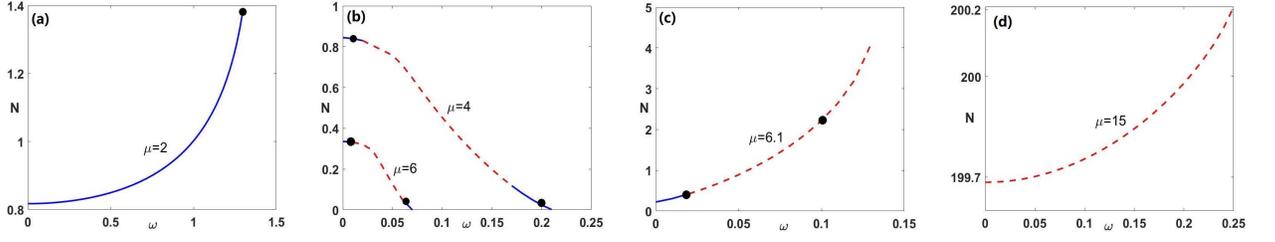}}}\hspace{-0.35in} \vspace{%
0.2in}
\caption{{\protect\small $N(\protect\omega )$ curves for droplets at
different values of }$\protect\mu $ {\protect\small (dashed: unstable;
solid: stable). (a) One-component droplets at $\protect\mu =2$, belonging to
the bottom branch in Fig. \protect\ref{stability}(A). (b) Two- and
three-component droplets at $\protect\mu =4$ and $\protect\mu =6$,
respectively (they belong to the bottom branch in Fig. \protect\ref%
{stability}(A)). (c) Four-components droplets at $\protect\mu =6.1$,
belonging to the bottom branch in Fig. \protect\ref{stability}(A). (d)
Two-component droplets at $\protect\mu =15$, belonging to the top branch in
Fig. \protect\ref{stability}(A).}}
\label{rotate_n}
\end{figure}

\begin{figure}[!t]
\centering
{\scalebox{0.75}[0.7]{\includegraphics{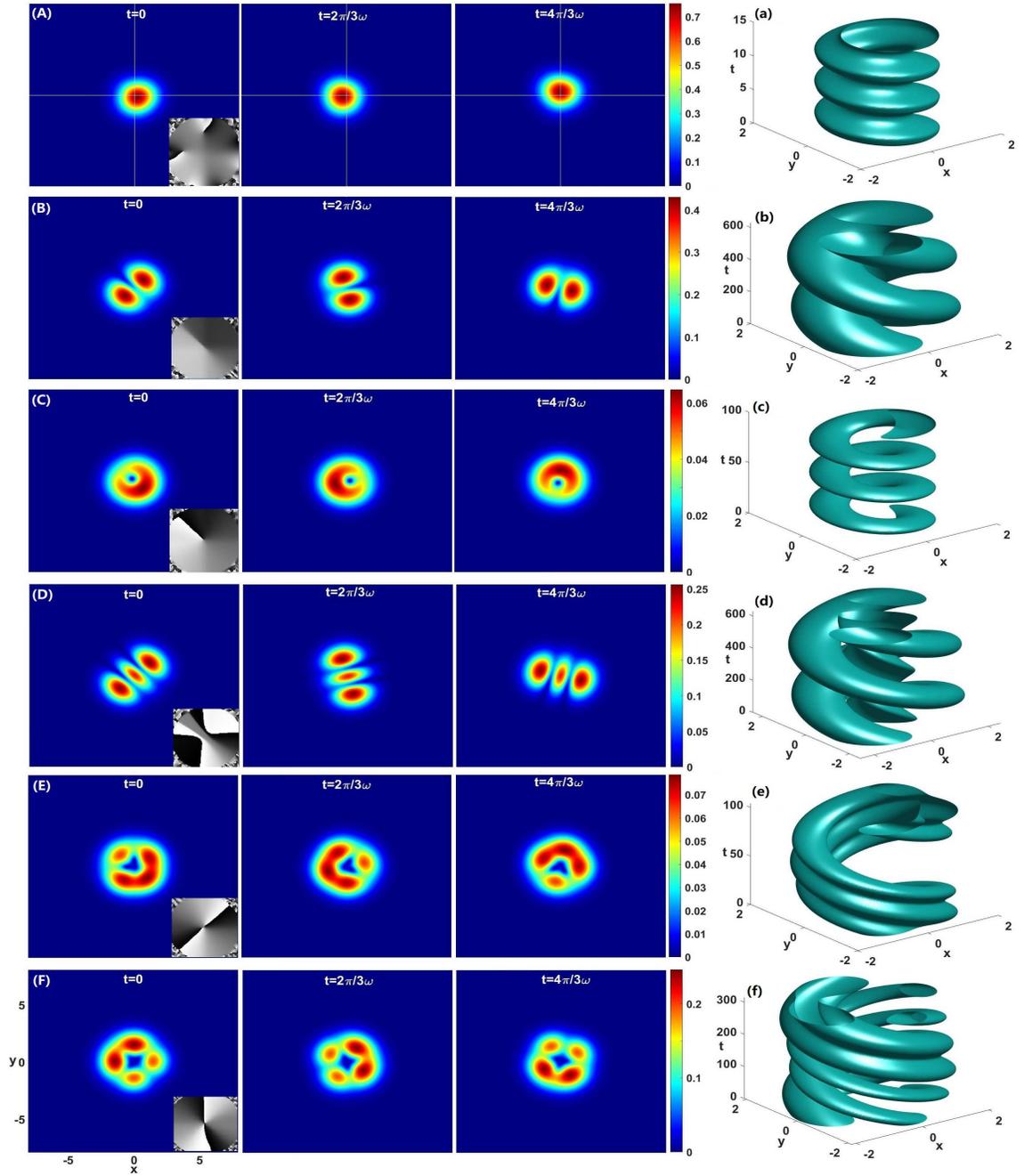}}}\hspace{-0.35in}
\vspace{0.1in}
\caption{{\protect\small Stable evolution of spinning asymmetric QDs,
displayed by means of density isosurfaces in the domain of $|x|,|\,y|\leq 8$%
. (A) A one-component droplet at $\protect\mu =2$, $\protect\omega =1.35$.
(B,C) Two-component droplets at $\protect\mu =4$, with $\protect\omega =0.01$
and $0.2$, respectively. (D,E) Three-component droplets at $\protect\mu =6$,
with $\protect\omega =0.01$ and $0.06$, respectively. (F) A four-component
droplet at $\protect\mu =6.1$, $\protect\omega =0.02$. (a-f) The
corresponding isosurface evolutions with the value of the contour lines
taken as the level of $(1/2)\max (|\protect\phi (\mathbf{r})|^{2})$. }}
\label{srotate}
\end{figure}
\vspace{0.1in} \textit{Droplets with the three- and four-component structures%
}.---$\mathcal{PT}$-symmetric QDs composed of three separated fragments
originate from the above-mentioned mode at $\lambda _{2}=6.1097$ which
corresponds to the second excited state of the linearized equation (\ref{3ds}%
). 
The respective QD family is represented by the rightmost curve $N(\mu )$ in
Fig. \ref{stability}(A), which exists at $\mu \geq 5.8187$. An example of
the density and phase structure produced by the numerical solution for the
stable two-component QD at $\mu =4$ (it corresponds to point \textbf{c1 }in
Fig. \ref{stability}(B)), is shown in Fig.~\ref{stability}(c1) for $\mu =6$,
whose phase structure reveals that it carries topological charge $m = 4$.
The increase of $N$ leads, as well as in the case of two-component modes, to
gradual attraction of the components, as shown for $\mu =8.2$ in Fig. \ref%
{stability}(c2), meanwhile the phase structure shows that the topological
charge becomes $m = 2$. At still larger values of $N$, such as the one
corresponding to $\mu =25$ in Fig. \ref{stability}(c3), the three components
fuse into a vortex octupole.

\vspace{0.1in} $\mathcal{PT}$-symmetric QDs composed of four separated
fragments originate from the above-mentioned mode at $\lambda _{3}=6.1739$,
which corresponds to the third excited state of the linearized equation (\ref%
{3ds}). 
The respective QD family is represented by the curve $N(\mu )$ in Fig. \ref%
{stability}(b3) labeled by points \textbf{d1, d2}, and \textbf{d3}. It
exists at $\mu \geq 5.8622$. Although this curve is close to the one
representing the two-component QDs, the two curves do not intersect (see
Fig.~\ref{stability}(B)). An example of the density and phase structure
produced by the numerical solution for the stable three-component QD at $\mu
=6.1$ (it corresponds to point \textbf{d1 }in Fig. \ref{stability}(B)), is
shown in Fig.~\ref{stability}(d1), whose phase structure implies that it
carries topological charge $m = 6$. The increase of $N$ again leads to
attraction between the components, as shown for $\mu =7.5$ in Fig. \ref%
{stability}(d2), which is shaped as a quadrupole droplet. At still larger
values of $N$, such as the one corresponding to $\mu =18$ in Fig. \ref%
{stability}(d3), the four components fuse pairwise to form a vortex dipole
containing two vortices and a phase at the center. The two remaining
separate components observed in the latter figure do not fuse with further
increase of $N$.

It may be relevant to consider additional species of $\mathcal{PT}$%
-symmetric QD modes which bifurcate from higher-order excited states of the
linearized equation (\ref{3ds}). This option will be elaborated elsewhere.

\section{2D spinning QDs and their stability}

\begin{figure}[t]
\centering
{\scalebox{0.8}[0.8]{\includegraphics{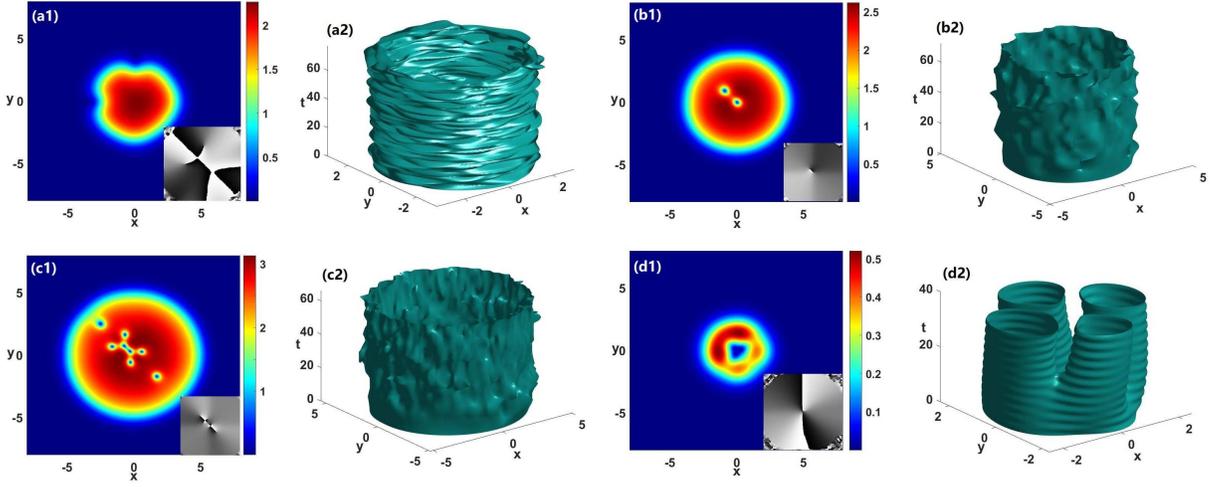}}}\hspace{-0.35in}
\vspace{0.1in}
\caption{{\protect\small Instable evolutions of spinning asymmetric QDs
displayed by means of density isosurfaces. (a1,a2) A one-component droplet
at $\protect\mu =9$, $\protect\omega =0.873$. (b1,b2) A two-component
droplet at $\protect\mu =15$, $\protect\omega =0.25$. (c1,c2) A
three-component droplet at $\protect\mu =25$, $\protect\omega =0.1$. (d1,d2)
A four-component droplet at $\protect\mu =6.1$, $\protect\omega =0.1$. }}
\label{unsrotate}
\end{figure}

In this section, we address QD modes produced by Eq. (\ref{3ds}), which
includes the spinning term with $\omega >0$. A family of symmetrically
shaped spinning QD solutions of the one-component type is characterized by
dependence $N(\omega )$ for a fixed chemical potential. For $\mu =2$, which
corresponds, at $\omega =0$, to the bottom branch of the $N(\mu )$
dependence in Fig. \ref{stability}(A), the $N(\omega )$ curve is displayed
in Fig.~\ref{rotate_n}(a). It is seen that the QD's norm monotonously
increases with the increase of the rotation frequency $\omega $. At $\omega
\approx 1.35$, the symmetric QD transforms into a slightly asymmetric
spinning one, and phase singularities originated from transverse boundary
and longitudinal boundary and gradually become larger, see Fig.~\ref{srotate}%
(A). And it is shown that the droplet drifts off the center. Further
increase of $\omega $ leads to disappearance of spinning droplets at the
critical value $\omega\approx 1.352$, but they remain stable as long as they
exist. For example, for $\omega =1.35$, stable rotation of the asymmetric
droplet and contour plots for the intensity, taken as the level of $%
(1/2)\max (|\phi (\mathbf{r})|^{2})$, are exhibited in Figs.~\ref{srotate}%
(A,a). Symmetric QDs belonging to the unstable segment of the $N(\mu )$
curve in Fig. \ref{stability}(A), such as one with $\mu =9$ (see Fig.~\ref%
{stability}(a2)) spontaneously transforms into an asymmetric one, which is
displayed in Fig.~\ref{unsrotate}(a1). However, the latter state turns out
to be unstable, as its norm is too large ($N\approx 91.53$), as shown in
Fig. \ref{unsrotate}(a2).

For QDs with a more sophisticated structure, which includes several
components, the dependences $N(\omega )$ for the modes with $\mu =4$ and $%
\mu =6$, belonging to the bottom branch in Fig. \ref{stability}(A), are
displayed in Fig.~\ref{rotate_n}(b).
At a critical value of $\omega $, the norm drops to zero, signaling that the
QD disappears at this point. In particular, at $\omega =0.01$ the two
components start to fuse and the phase changes, see Fig.~\ref{srotate}(B).
At $\omega =0.17$, the two components are completely fused into an
asymmetric spinning droplet, as seen in Fig.~\ref{srotate}(C). However, the
spinning QDs are unstable in the region of $0.01<\omega \leq 0.17)$.
Similarly, a three-component QD with $\mu =6$ tend to fuse, following the
increase of $\omega $. However, as shown in Fig. \ref{srotate}(E), a novel
QD appears with four-density peaks at $\omega =0.06$, which is a mode
connecting different states. In the regions $[0\leq \omega \leq 0.01]$ and $%
[0.06\leq \omega \leq 0.07]$ the spinning droplets are stable (see Figs.~\ref%
{srotate}(D,d,E,e)). However, vortex QDs belonging to the top branch in Fig. %
\ref{stability}(A) are, generally, unstable. For example, the $N(\omega )$
curve for $\mu =15$ is displayed in Fig.~\ref{rotate_n}(d). In Fig. \ref%
{unsrotate}(b2), the rotation drives the two vortices to the top left in the
form of a vortex dipole at $\mu =15$, $\omega =0.25$. Similarly, in Figs. %
\ref{unsrotate}(c1,c2), for $\omega =0.1$, $\mu =25$ the vortices form an
unstable vortex octupole.

For symmetric four-component QDs, the $N(\omega )$ dependence for $\mu =6.1$%
, belonging to the bottom branch in Fig. \ref{stability}(A), is displayed in
Fig. \ref{rotate_n}(c). In this case, the dependence is a monotonously
growing one, switching from stability to instability. For instance, at $%
\omega =0.02$, the four components start to fuse pairwise, and the density
grows in two of them, as shown in Fig.~\ref{srotate}(F). The QDs become
unstable with the further increase of $\omega $, see Figs.~\ref{unsrotate}%
(d1,d2).

\begin{figure}[t]
\centering
{\scalebox{0.8}[0.8]{\includegraphics{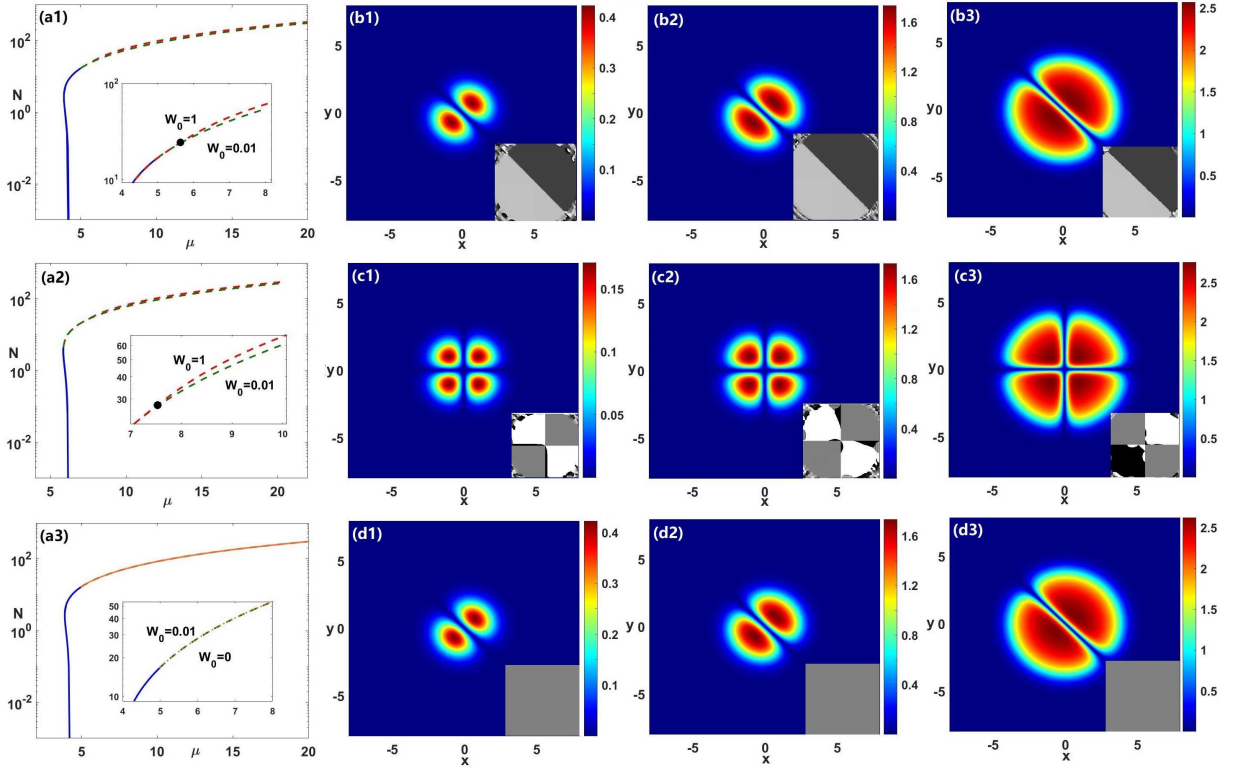}}}\hspace{-0.35in}
\vspace{0.1in}
\caption{{\protect\small (a1) Dependences of norm $N$ on chemical potential $%
\protect\mu $ for two-component nonspinning ($\omega=0$) QDs at $W_{0}=1$ and $W_{0}=0.01$. (a2)
Dependences of $N$ on chemical potential $\protect\mu $ for four-component
QDs at $W_{0}=1$ and $W_{0}=0.01$. (a3) Dependences of norm $N$ on chemical
potential $\protect\mu $ for two-component QDs at $W_{0}=0$ and $W_{0}=0.01$
(dashed: unstable; solid: stable). Dipole droplets at $W_{0}=0.01$: (b1)
$\protect\mu =4$, (b2) $\protect\mu =5.75$, (b3) $\protect\mu =15$.
Quadrupole droplets at $W_{0}=0.01$: (c1) $\protect\mu =6.1$, (c2)  $%
\protect\mu =7.5$, (c3) $\protect\mu =18$. Dipole droplets at $W_{0}=0$  (real potential):
(d1)  $\protect\mu =4$, (d2) $\protect\mu =5.75$, (d3)   $\protect%
\mu =15$.}}
\label{01unrotate}
\end{figure}

\begin{figure}[t]
\centering
{\scalebox{0.8}[0.8]{\includegraphics{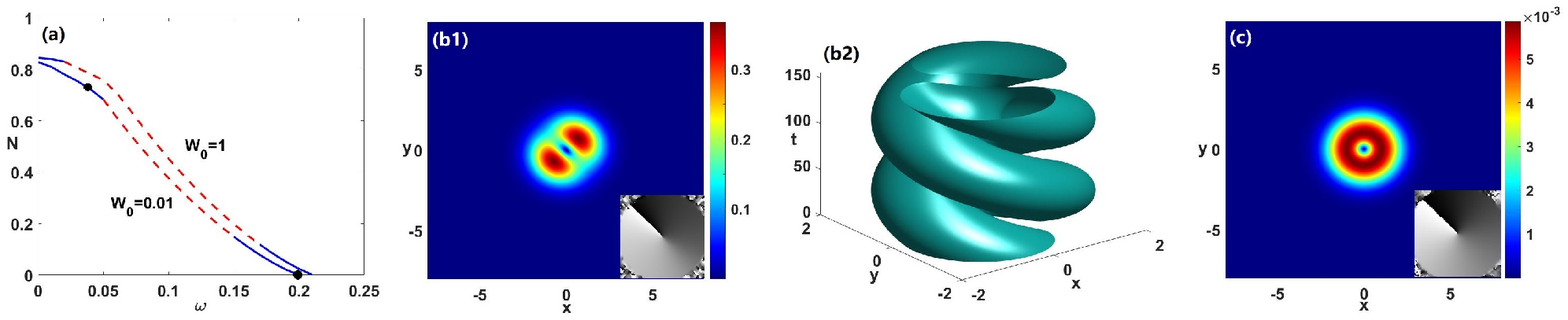}}}\hspace{-0.35in} \vspace{%
0.1in}
\caption{{\protect\small (a) $N(\protect\omega )$ curves for QDs with $%
\protect\mu =4$ and $W_{0}=1$ or $W_{0}=0.01$ (dashed: unstable; solid:
stable). (b1,b2) The evolution of a spinning droplet at $\protect\omega %
=0.04 $ is displayed by means of the density isosurface. (c) A spinning
droplet at $\protect\omega =0.2$. }}
\label{01rotate}
\end{figure}

\section{Effect of $\mathcal{PT}$-symmetric potential on the (non)spinning QDs}

\begin{figure}[t]
\centering
{\scalebox{0.6}[0.6]{\includegraphics{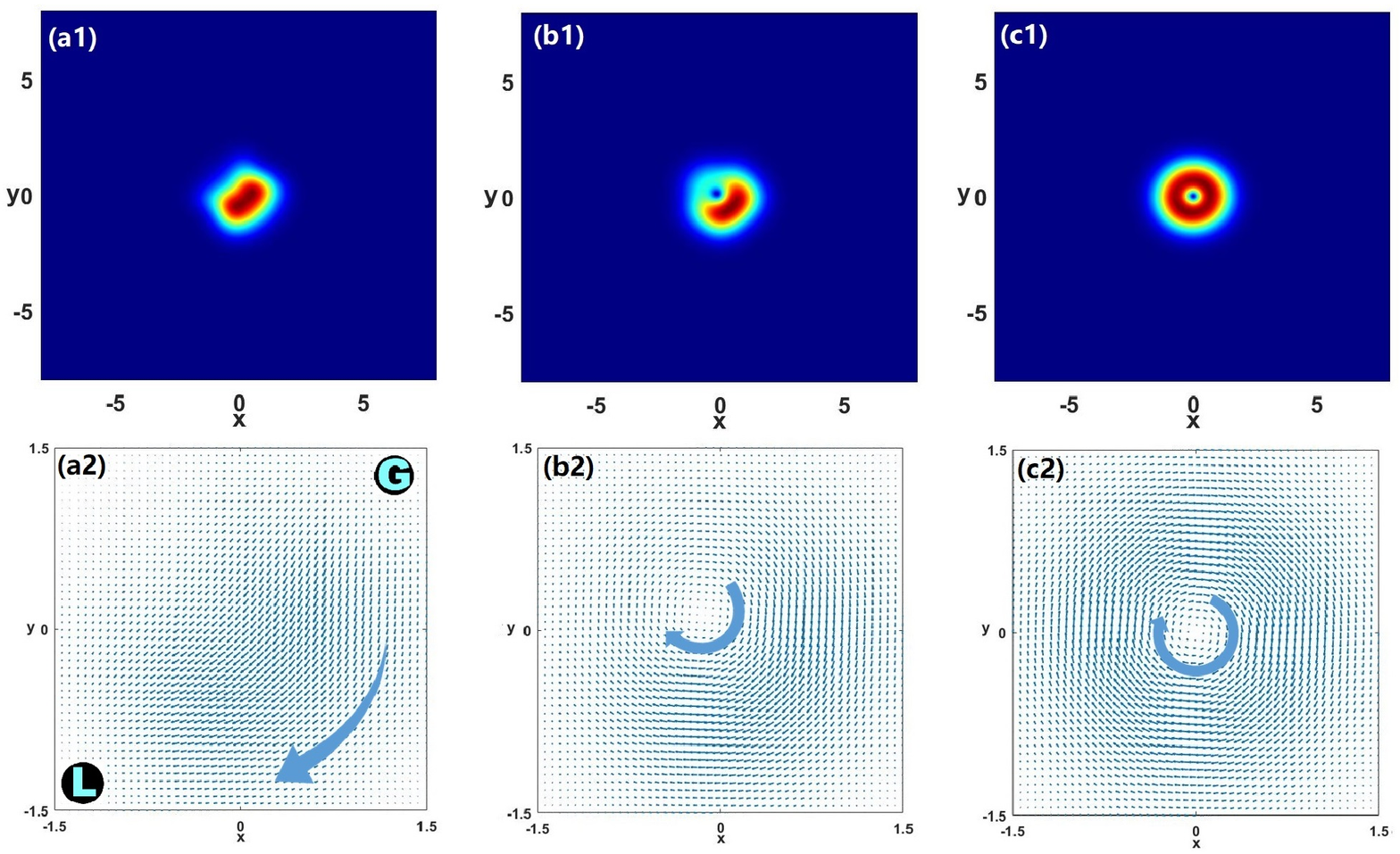}}}\hspace{-0.35in} \vspace{%
0.1in}
\caption{{\protect\small The absolute value $|\vec{S}|$ of the Poynting
vector and power flow at $\protect\mu =4$: (a1,a2) $W_{0}=1$, $\protect%
\omega =0.01$, with points $\mathrm{G}$} and{\protect\small \ $\mathrm{L}$
points denoting regions where the gain and loss are concentrated, as per Eq.
(\protect\ref{hg}). (b1,b2) $W_{0}=1$, $\protect\omega =0.2$; (c1,c2) $%
W_{0}=0.01$, $\protect\omega =0.04$. }}
\label{pf}
\end{figure}

To explore the effect of $\mathcal{PT}$-symmetric potential (\ref{hg}) on
the (non)spinning QDs, we vary the strength, $W_{0}$, of gain-loss distribution
(imaginary part of the potential).

\subsection{The nonspinning case}

For the nonspinning case, fixing other potential parameters, $V_{0}$
and $V_{1}$, for QDs with two components we compare dependencies of norm $N$
on chemical potential $\mu $ for $W_{0}=1$ and $W_{0}=0.01$ in Fig.~\ref%
{01unrotate}(a1). As $\mu \prec 5.75$, The norms $N$ are little affected by gain-loss
distribution $W(\mathbf{r})$ (see, for example, Fig.~\ref{stability}(b1) at $W_0=1$
and Fig.~\ref{01unrotate}(b1) at $W_0=0.01$ for $\mu=4<5.75$).
However at $\mu \succeq 5.75$ the QD's two components start to fuse at $W_{0}=1$ (see Fig.~\ref{stability}%
(b2)), while for $W_{0}=0.01$ the components of the dipole QD are still well
separated, and it remains broad, see Figs. \ref{01unrotate}(b2, b3). 
To clarify the role of the $\mathcal{PT}$-symmetric potential, we further
change the strength, $W_{0}$, of the gain-loss distribution. For QDs with
two components, the dependencies of norm $N$ on chemical potential $\mu $
for $W_{0}=0.01$ and $W_{0}=0$ (the real potential)\ are shown in Fig.~\ref%
{01unrotate}(a3). It can be seen that the curves corresponding to $W_{0}=0.01
$ and $0$ are virtually identical. Profiles of droplets at $W_0=0.01, 0$ are almost the same for
equal values of chemical potential $\mu $ by comparing Figs.~\ref{01unrotate}(b1-b3) for $W_0=0.01$ and
 Figs.~\ref{01unrotate}(d1-d3) for $W_0=0$, respectively. Similarly, for four-component QDs the curves corresponding to $%
W_{0}=1$ and $0.01$ are nearly identical at small values of $N$ (i.e., at $\mu \prec  7.5$)
(see, for example, Fig.~\ref{01unrotate}(a2), and compare Fig.~\ref{stability}(d1) at $W_0=1$ and Fig.~\ref{01unrotate}(c1) at $W_0=0.01$ for $\mu=6.1<7.5$). However, at larger $N$ (i.e., at $\mu \succeq 7.5$) the two curves split apart,
see Fig.~\ref{01unrotate}(a2). For $W_{0}=1$, the four-component QDs start to fuse in
pairs (see Figs.~\ref{stability}(d2, d3)), while each component gets broader for $W_{0}=0.01$, see Figs.~\ref%
{01unrotate}(c2,c3).

\subsection{The spinning case}

In the spinning regime ($\omega >0$), the gain-loss distribution $W(\mathbf{r%
})$ affects the symmetry of the droplets. For example, $N(\omega )$ curves
for QDs with $\mu =4$ at $W_{0}=1$ and $W_{0}=0.01$ are displayed in Fig.~%
\ref{01rotate}(a). With the increase of $\omega $, the norm of the two QD
families drops to zero. As shown above, at $W_{0}=1$ the two QD components
start to fuse, ending up by forming an asymmetric droplet in Figs.~\ref%
{srotate}(B,C). However, when the strength of the gain-loss distribution is
small, the two QD components begin to merge symmetrically, eventually
forming a symmetric vortex droplet in Figs.~\ref{01rotate}(b1,c)). They are
stable in the regions $0\leq \omega \leq 0.05$ and $0.15\leq \omega \leq 0.2$%
.

Due to the presence of the imaginary part of the potential, QDs are
amplified (absorbed) in gain (loss) regions. To outline the structure of the
nonlinear modes, we use the Poynting vector, which determines the density of
the power flow from gain to loss:
\begin{equation}
\begin{array}{rl}
\vec{S}(x,y)=\frac{i}{2}(\phi \nabla \phi ^{\ast }-\phi ^{\ast }\nabla \phi
), &
\end{array}
\label{S}
\end{equation}%
where $\phi $ is the nonlinear localized mode produced by Eq. (\ref{3ds}).
In the nonspinning regime, the power always flows from the gain to the loss.
As the angular velocity $\omega $ increases, the direction of the power flow
shifts, and then a vortex is formed, as seen in Figs.~\ref{pf}(a1,a2,b1,b2).
The strong gain-loss distribution makes the QDs become asymmetric, see Figs.~%
\ref{srotate}(B,C). On the other hand, if the gain-loss term is weak with
small $W_{0}$, the effect of the rotation on the energy flow is dominant.
Therefore, in that case the energy flow supports symmetric QDs, see Figs.~%
\ref{pf}(c1,c2).

\begin{figure}[!t]
\centering
{\scalebox{0.8}[0.8]{\includegraphics{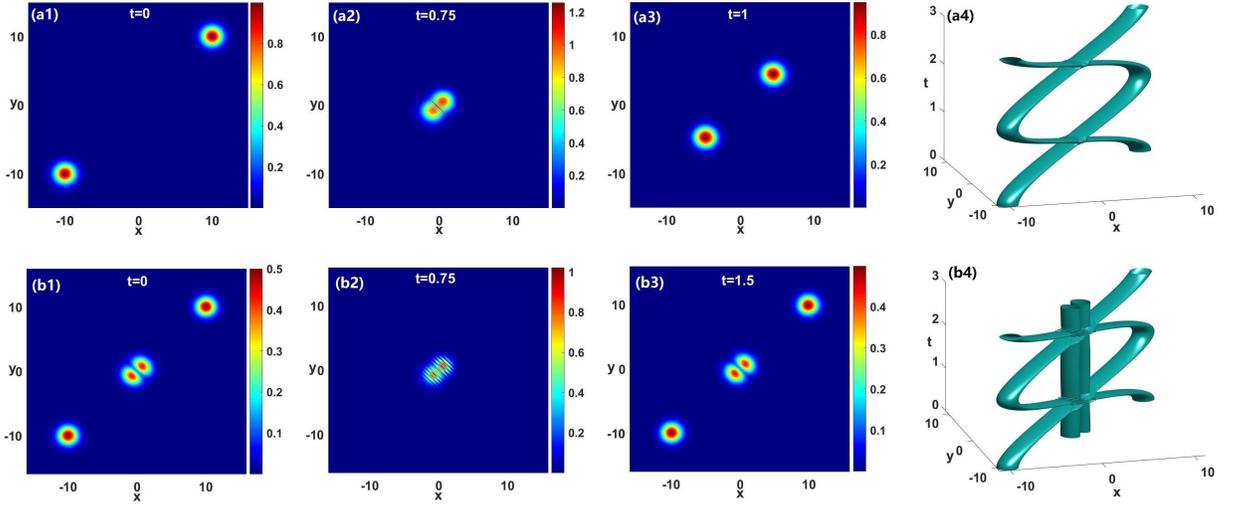}}}\hspace{-0.35in} \vspace{%
0.15in}
\caption{{\protect\small (a1-a4) The collision of two nonspinning droplets
(with $A=1$ and $B=0$ in Eq. (\protect\ref{coll})) at different times, and
the respective isosurface evolution. (b1-b4) The collision of two
nonspinning droplets (with $A=0.5$ and $B=1$) at different times and the
respective isosurface evolution, where $\protect\phi (\mathbf{r})$ in Eq. (%
\protect\ref{coll}) is the dipole droplet at $\protect\mu =4$. The strength
of the gain-loss distribution is $W_{0}=1$. }}
\label{collision}
\end{figure}

\section{Interactions between spinning or nonspinning QDs}

For collisions between two or three QDs, we address changes of the state of
spinning or nonspinning droplets produced by the collisions. The corresponding initial
condition is taken as a superposition of separated droplets:
\begin{equation}
\psi (\mathbf{r},t=0)=A[\phi _{1}(\mathbf{r}-\mathbf{r}_{0})+\phi _{1}(%
\mathbf{r}+\mathbf{r}_{0})]+B\phi (\mathbf{r}),  \label{coll}
\end{equation}%
where $\phi _{1}(\mathbf{r})$ is the exact QD solution given by Eq.~(\ref%
{sol2d}), and $\phi (\mathbf{r})$ may be an additional solution of
stationary equation (\ref{3ds}). Parameters $\mathbf{r}_{0}=(x_{0},y_{0})$
define initial positions of the colliding QDs, and coefficients $A$ and $B$
determine the composition of the input.

\begin{figure}[t]
\centering
{\scalebox{0.8}[0.8]{\includegraphics{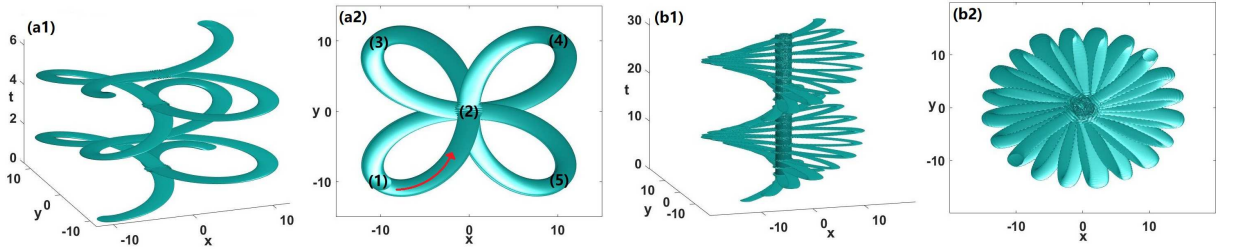}}}\hspace{-0.35in} \vspace{%
0.15in}
\caption{{\protect\small (a1) The isosurface evolution for the collision
of two spinning droplets with $\protect\omega =1$, and (a2) its projection onto the }$\left(
{\protect\small x,y}\right) ${\protect\small \ plane, corresponding to nonspinning ones
 in Fig.~\protect\ref{collision}(a4).
(b1) The isosurface
evolution for the collision of three spinning droplets with $\protect\omega =0.2$, and its
projection onto the }$\left( {\protect\small x,y}\right) ${\protect\small \
plane, corresponding to  nonspinning ones in Fig.~\protect\ref{collision} (b4).  The strength of the gain-loss distribution is $W_{0}=1$. }}
\label{collision1}
\end{figure}

The real part of potential (\ref{hg}) draws the QDs towards the center and
leads to the collision. First, we consider the collision of a pair of
nonspinning droplets ($\omega=0$), which corresponds to $A=1$ and $B=0$ in Eq. (\ref{coll}%
). Figures~\ref{collision}(a1, a3, a4) demonstrate an elastic collision, which
preserves shapes of both droplets. At the collision point, at $t=0.75$, an
interference pattern appears, which is a distinctive feature of the
interplay of coherent matter waves, as seen in Fig.~\ref{collision}(a2).
For the collision of two spinning QDs, for example, $\omega =1$,
Figure~\ref{collision1}(a1) demonstrates an elastic collision between two spinning QDs,
by means of the density isosurface taken at $(1/2)\max (|\phi (\mathbf{r})|^{2})$.
Figure~\ref{collision1}(a2) additionally shows the evolution of the
isosurface in projection onto the $\left( x,y\right) $ plane in the course
of the rotation period $2\pi /\omega $. This plot shows that $\phi _{1}(%
\mathbf{r}+\mathbf{r}_{0})$ follows trajectory $(1)\rightarrow
(2)\rightarrow (3)\rightarrow (2)\rightarrow (4)\rightarrow (2)\rightarrow
(5)\rightarrow (2)\rightarrow (1)$, finally returning to the initial
position.

For the collision of three nonspinning droplets in Eq.~(\ref{coll})
corresponding to $\mu =4$, with $A=0.5$ and $B=1$, where $\phi (\mathbf{r})$
is the respective dipole droplet, the collision remains
perfectly elastic, with the interference pattern appearing in the course of
the collision, see Figs. \ref{collision}(b1)-(b4).
For the collision of thee spinning QDs, for example, $\omega =0.2$,
Fig.~\ref{collision1}(b1) also demonstrates an elastic collision between three spinning QDs,
by means of the density isosurface taken at $(1/2)\max (|\phi (\mathbf{r})|^{2})$. The interaction seems to be
complicated (see Fig.~\ref{collision1}(b2) for the projection of the isosurface onto the $\left( x,y\right) $ plane).

These results indicate that collisions between 2D $\mathcal{PT}$-symmetric spinning or nonspinning
QDs are similar to what is known about other nonintegrable models, such the
one for one-dimensional QDs \cite{1d-2,songqd}.

\section{Conclusions and discussions}

We have introduced the 2D dynamical model for the BEC under the action of
the $\mathcal{PT}$-symmetric potential, including the LHY correction. The
linear spectrum of the model remains real up to a critical strength of the
imaginary part of the potential. In the region of the unbroken $\mathcal{PT}$
symmetry, several families of QDs (quantum droplets) originating from the
linear modes are obtained, in the form of multipolar modes with smaller
norms, and vortex ones with larger norms. A particular solution is obtained
in the exact analytical form. Stability of these QD states is explored by
means of numerical methods. The effect of the rotation (spinning) on the QDs
is studied too, showing that the asymmetric energy flow, resulting from the
imbalance of the gain-loss distribution and rotation, creates asymmetric
spinning QDs. Collisions between spinning QDs are explored too, with a
conclusion that they tend to be elastic.

The analysis presented in this paper can be extended for solitons, including
spinning ones, in models combining the nonlinearity and $\mathcal{PT}$
symmetry with fractional diffraction, cf. Refs. \cite{PT1,PT2}.

\vspace{0.1in}\noindent\textbf{Acknowledgments} \vspace{0.1in}

The work was supported by the National Natural Science Foundation of China
(No. 11925108) and Israel Science Foundation (No. 1695/2022).




\end{document}